\documentclass[aps,prd,preprint,superscriptaddress,nofootinbib]{revtex4-1}

\usepackage{latexsym}
\usepackage{amsmath}
\usepackage{amssymb}
\usepackage{amsfonts}
\usepackage{bm}
\usepackage{mathtools}
\usepackage{braket}

\usepackage{color}
\definecolor{purple}{rgb}{0.5,0,0.5}
\definecolor{blue}{rgb}{0.0,0,0.9}
\definecolor{prdblue}{rgb}{0.133,0.118,0.498}
\usepackage[colorlinks=true, pdfstartview=FitV, linkcolor=prdblue, citecolor= prdblue, urlcolor=prdblue]{hyperref}

\usepackage{supertabular} 
\usepackage{placeins}
\usepackage{epsfig}
\usepackage{graphicx}

\newcommand{\RNum}[1]{\uppercase\expandafter{\romannumeral #1\relax}}

\begin{document}


\title{Quantum dynamics for massless particles in Brinkmann spacetimes}


\author{\'Alvaro Duenas-Vidal}
\email[]{alvaro.duenas@epn.edu.ec}
\affiliation{Departamento de Física, Escuela Politécnica Nacional, Quito 170143, Ecuador}

\author{Jorge Segovia}
\email[]{jsegovia@upo.es}
\affiliation{Departamento de Sistemas F\'isicos, Qu\'imicos y Naturales, \\ Universidad Pablo de Olavide, E-41013 Sevilla, Spain}


\date{\today}

\begin{abstract}

In Classical Dynamics, Eisenhart lift connects the dynamics of  null geodesics in a Brinkmann spacetime with a continuous family of Hamiltonian systems by means of a suitable projection. In this work we explore the possibility of building a model for quantum dynamics of massless particles propagating inside a Brinkmann spacetime from the Einsenhart lift. As a result, we describe spatial tunneling between regions classically disconnected for certain class of null geodesics  because of curvature.  Also we describe entangled states arising from observers who have a limited access to the whole Brinkmann space. Finally we explore the possibility to find a quantum field theory behind these quantum phenomena.    

\end{abstract}


\maketitle


\section{Introduction}\label{sec_1}

In a microscopic scale, mass particles in a scalar potential evolve showing a collection of phenomena characteristic of Quantum Mechanics  as, for example, quantum tunneling, barrier penetration, etc.  As a particular example we can choose the scalar potential of Newtonian gravity and, therefore, we would expect that the microscopic dynamics of a mass particle subjected  to gravity follows the rules of Quantum Mechanics as a response to the Newtonian gravitational potential. However,  gravity is driven by curvature, no by means of a scalar potential. On the other hand, in the framework of General Relativity, massless particles follow null geodesics that see curvature. So a natural question arises: do massless particles response to spacetime curvature, in a microscopic scale, following the rules of Quantum Mechanics? Note that we are taking about a scale in which particles behave following Quantum Mechanics but gravity is still described by classical (\emph{i.e.} no quantum) General Relativity. \\

The role of curvature in Quantum Mechanics has been considered many times in literature. Early works explored the generalization of Quantum Mechanics in the Euclidean space to Riemannian manifolds. In this way, in \cite{PhysRevA.23.1982} the role of the  curvature was computing through the constrain of the dynamics to curved submanifolds of the Euclidean space. On the other hand, in \cite{RevModPhys.29.377} the curvature was incorporated in a natural way by means of a connection in the fiber bundle of complex number over the space. Each model brings to different results and problems,  and nowadays in not clear what is the correct answer. With respect to the modeling of the quantum dynamics for geodesics in curved spacetimes, it has been addressed both in the framework of Quantum Geometry \cite{Beggs:2022bhw, Liu:2021mgs}, where non-commutative spaces serve as play model for Quantum Gravity, and in the context of quantum correcting to gravity \cite{Miller:2016gwd, Deng:2020yfm,Gwynne:2019der, Gwynne:2020nex, Emelyanov:2020wxr, Dalvit:1999wd}. Finally, issues concerning if gravity sees quantum physics was considered in \cite{Colella:1975dq} and, more recently, in \cite{Krisnanda:2019glc, Carlesso:2019cuh}.\\

In this work we propose an approach different of the ones previously mentioned. Here we do not pretend to develop any quantum model for gravity either to introduce quantum corrections to gravity. We neither build a covariant formulation of Quantum Mechanics. Our point of view is to move in a scale where gravity is well described by General Relativity  but particles exhibit quantum dynamics, and describe this quantum dynamics for a selected set of observers. For achieving this goal, our point of departure is the Eisenhart lift.\\

The Eisenhart lift arises in Classical Dynamics as an attempt to \emph{geometrize} the dynamics of Hamiltonian systems \cite{Cariglia:2015bla, Cariglia:2016oft}. Being more specific, the Eisenhart mechanism works \emph{lifting} the dynamics of some Hamiltonian system to a higher dimensional space by introducing two addition coordinates. The key point is that the lifted trajectories  correspond to  null geodesics in a pseudo-Riemannian space with Lorentzian signature, such that the coupling constant to scalara and vector potentials appearing in the original Hamiltonian system corresponds to a conserved quantity over null geodesics. This conserved quantity is related to a covanriantly conserved Killing vector and, thus, the null geodesics are in a Brinkmann spacetime \cite{Brinkmann:1925fr, stephani_2003, Blau:2003dz, Blau:2004yi}. As a consequence, the inverse of the Eisenhart lift gives a projection of the dynamics of null geodesics in a Brinkamnn space over a continuous uniparametric family of Hamiltonian systems depending on the coupling constant.\\ 

Based on the connection previously mentioned between null geodesics and a continuous family of Hamiltonian systems, in this paper we propose to define quantum dynamics for massless particles in Brinkmann spacetimes by \emph{quantizing} the dynamics in each one of the Hamiltonian systems of the continuous family served by the Eisenhart lift. In some way this is not a new idea, since second quantization in Quantum Field Theory is carried out by decomposing the field in a continuum of modes and, then, describing each mode by means of the quantum harmonic oscillator. \\

The paper is organized as follow: in Sec.~\ref{sec_2} we discuss briefly the Eisenhart lift and its relation with null geodesics in Brinkmann spacetimes, setting the connection between null geodesics and a continuous family of \emph{reduced} Hamiltonian systems. In Sec.~\ref{sec_3}, as a warm up exercise,  we pay some attention to the Eisenhart lift of autonomous and natural Hamiltonian systems. In Sec.~\ref{sec_4} we define quantum dynamics for massless particles in Brinkmann spacetimes from this continuous family of reduced Hamiltonian systems, doing an special emphasis over stationary Brinkamnn spacetimes. In Sec.~\ref{sec_5} we extract some consequences from  the model previously developed, and  we describe quantum tunneling and entanglement for null particles in Brinkmann spacetimes. Finally, in Sec.~\ref{sec_6}, the possibility to find a Quantum Field Theory behind the quantum-mechanical model  developed is explored, while in Sec.~\ref{sec_7} some conclusions and future work are addressed.

\section{Eisenhart lift and Brinkmann spaces} \label{sec_2}

The Eisenhart lift connects the dynamics of Hamiltonian systems with the dynamics of null particles in a pseudo-Riemannian manifold. In this section we show briefly how this is done and the connection with Brinkmann spaces. Deeper details in the Eisenhart lift can be found in Refs. \cite{Cariglia:2015bla, Cariglia:2016oft}  while, for Brinkamnn spaces,  canonical texts as Refs. \cite{Ehlers:1962zz, stephani_2003}  can serve as a complete view of the subject.

\subsection{The Eisenhart lift briefly}

Let $(F, \omega, H(\lambda) )$ be a Hamiltonian system, where $\omega = \sum_i dq^i \wedge dp_i$,  $i = 1, \ldots n$, $F = T^* Q_n$ and $\lambda$ the coupling constant to external potentials. The Eisenhart lift  consists in building a new Hamiltonian system $(\mathcal F, \Omega, \mathcal H)$, with $\Omega = \omega + du\wedge dp_u + dv \wedge p_v $, $\mathcal F = T^*\mathcal Q_{n+2}$ and, 
\begin{equation}\label{LiftedHamiltonian}
\mathcal H = \frac{1}{2} \bar {g}^{ij} \left(p_i - \frac{p_v}{2} \mathcal A_i\right) \left(p_j - \frac{p_v}{2} \mathcal A_j\right) - \frac{p_v^2}{2}\,\Phi +  p_v p_u, 
\end{equation}
such that, 
\begin{equation}\label{LambdaSectorHamiltonian}
\mathcal H_\lambda = \left.H(\lambda)\right|_{t \rightarrow  u} +  \lambda p_u, 
\end{equation}
where the the subscript $\lambda$ means 'valued over $p_v = \lambda$'. In the most general case, $( F, \omega, H(\lambda))$ is a no-autonomous and no-natural Hamiltonian system with scalar potential $\varphi(t,q)$, vector potential $A_i(t,q)$ and kinetic metric $g_{ij}(t,q)$ given by, 
\begin{equation}\label{liftedpotentials}
 g_{ij} = \left.\bar{g}_{ij}\right|_{u\rightarrow t},\quad \varphi = -\frac{1}{2} \left.\Phi\right|_{u\rightarrow t}, \quad  A_i = \frac{1}{2}\left.\mathcal A_i\right|_{u \rightarrow t}.
\end{equation}
 We impose the condition that $det(g) \ne 0$ for any value of $t$ and $q^i$. Then, positiveness of the kinetic term fixes $g_{ij}$ as a uniparametric family of Riemannian metrics over the configuration space $Q_n$ for each value of $t$. From now on we shall call $(\mathcal F, \Omega, \mathcal H)$ the \emph{lifted} Hamiltonian system, while we shall call $( F, \omega, H(\lambda))$ the \emph{reduced} Hamiltonian system with coupling $\lambda$. \\

The "magic" of the Eisenhart lift is that the lifted Hamiltonian covers the dynamics of the reduced Hamiltonian. For showing this fact, note that $v$ is a cyclic coordinate in the lifted Hamiltonian Eq.~\eqref{LiftedHamiltonian} and, therefore, there are curves from the lifted Hamiltonian flow that satisfies $p_v = \lambda$. Then, 
\begin{equation}\label{equationsOfMotion}
\frac{\partial \mathcal H_\lambda}{\partial q^i} = \left.\frac{\partial H(\lambda)}{\partial q^i}\right|_{t\rightarrow u}, \quad \frac{\partial \mathcal H_\lambda}{\partial p_i} = \left.\frac{\partial H(\lambda)}{\partial p_i}\right|_{t\rightarrow u}, 
\end{equation}
and the Hamiltonian system $(\mathcal F, \hat \Omega, \mathcal H )$ includes the dynamics of  $(F, \omega, H(\lambda))$ for each value of the coupling $\lambda$. For $p_v = \lambda$, the set of equations of motion is completed with, 
\begin{equation}\label{CompleteEqsOfMotion}
\begin{split}
\dot u &= \lambda, \\
\dot{v} &= - \frac{1}{4}\bar g_{ij}  \mathcal A_i (2p_j - \lambda \mathcal A_j) - \lambda \Phi + p_u, \\
\dot{p}_u &= -\frac{1}{8}\partial_u \bar g_{ij}\, (2p_i - \lambda \mathcal A_i)(2p_j - \lambda \mathcal A_j) + \frac{\lambda}{4}\bar g_{ij}\, \partial_u \mathcal A_i\, (2 p_j - \lambda \mathcal A_j) + \frac{\lambda^2}{2} \partial_u \Phi, \\
\dot{p_v} &= 0. 
\end{split}
\end{equation}
Note that the value of the momentum  $p_v = \lambda$ gives the coupling of the reduced system to the scalar and vector potentials Eq.~\eqref{liftedpotentials} and, so, the coupling becomes dynamical in the lifted system.\\

Obviously, not all curves of the Hamiltonian flow in $(\mathcal F, \Omega, \mathcal H)$ projects over the Hamiltonian flow of the reduced system $(F, \omega,  H(\lambda_0))$ for some fixed coupling $\lambda_0$, since we have to choose the curves with $p_y = \lambda_0$. In a mathematical language, defining $\hat{\Gamma}_{\lambda_0}$ as the set of curves in $\mathcal Q_{n+2}$ solving the lifted equations of motion for $p_v = \lambda_0$,  the projection, 
\begin{equation}\label{EisenhartProjection}
\pi: \mathcal Q_{n+2} \rightarrow  Q_{n}, \quad \pi\left(q^1, \ldots q^n, u, v \right) = (q^1, \ldots q^n),
\end{equation}
induces a map, 
\begin{equation}\label{EpiyectiveMap}
\hat \pi : \hat{\Gamma}_{\lambda_0} \rightarrow \Gamma, 
\end{equation}
where $\Gamma$ is the set of curves in $Q_n$ solving the equations of motion in $(F, \omega, H(\lambda_0) )$. Even under the restriction to the sector $p_v = \lambda_0$ of the lifted dynamics, the map Eq.~\eqref{EisenhartProjection} is not  a one-to-one map. Since the $u$ coordinate replaces the time in  the original Hamiltonian  system, the  lifted Hamiltonian is autonomous and, therefore, 
\begin{equation}\label{epsilonfromGeodesics}
\varepsilon_\gamma =\mathcal H|_{\gamma(t)}, 
\end{equation}
for any curve $\gamma(t) \in \hat{\Gamma}_{\lambda_0}$, is a conserved quantity. This conserved quantity  does not arise from the dynamics in the configuration space $Q_n$ and, therefore, it is a free parameter. Thus, the inverse of  the map Eq.~\eqref{EpiyectiveMap} gives and infinity set of curves in $\hat{\Gamma}_{\lambda_0}$ for each curve in $\Gamma$, one curve for each possible value of $\varepsilon_\gamma$. Fixing a value $\varepsilon_\gamma$ gives a one-to-one map. For simplicity we decide to fix $\varepsilon_\gamma = 0$ and, then, we get the bijection:
\begin{equation}\label{oneToOneMap}
\hat{\pi} : \hat{\Gamma}^0_{\lambda_0} \simeq \Gamma, 
\end{equation}
where $\hat{\Gamma}^0_{\lambda_0}$ is the set of curves in $\mathcal Q_{n+2}$ solving the lifted equations of motion with $p_v = \lambda_0$ and $\varepsilon_\gamma = 0$.\\ 

The power of the Eisenhart lift yields over the fact that the lifted Hamiltonian is homogeneous of second degree in momenta. As a consequence the lifted Lagrangian formally coincides with the lifted Hamiltonian, and a short computation gives, 
\begin{equation}\label{liftedLagrangian}
\mathcal L = \frac{1}{2} \bar{g}_{ij}\,\dot q^i \dot q^j + \frac{\Phi}{2}  \dot u^2 +  \dot u\left( \dot v + \frac{1}{2}   \mathcal A_i  \dot q^i\right).
\end{equation}
This Lagrangian can be re-interpreted as the one for affine-parametrized  geodesics in $\mathcal Q_{n+2}$ with metric, 
\begin{equation}\label{BrinkmannMetric}
ds^2 = 2dvdu  +  \left(  \Phi(u,q) du + \mathcal A_i(u,q) dq^i \right) du + \bar{g}_{ij}(u,q) dq^i dq^j. 
\end{equation}
In addition, a simple computation gives, 
\begin{equation}\label{detG}
\mbox{det}(G) = - 4 \ \mbox{det}(\bar{g}), 
\end{equation}
and, thus, $G_{ab}$ equips $\mathcal Q_{n+2}$ with a pseudo-Riemannian geometry with Lorentzian signature.  In this way we gets an effective \emph{geometrization} of the classical dynamics of the set of  reduced Hamiltonian systems $( F, \omega, H(\lambda))$, $\lambda \in \mathbb R$,  by means of the Eisenhart lift. Since the lifted Lagrangian  coincides formally with the lifted Hamiltonian, we have, 
\begin{equation}\label{liftedEnergyFromGeodesics}
\varepsilon_\gamma = \mathcal L(\gamma, \dot \gamma) = \dot \gamma^a \dot \gamma^b G_{ab}. 
\end{equation}
On the other hand, the vector field $\vec{l} = \partial_v$ is a Killing field of the line element Eq.~\eqref{BrinkmannMetric} and, 
\begin{equation}\label{PyFromGravity}
l^a \dot \gamma^b G_{ab} = p_v. 
\end{equation}
Thus, 
\begin{equation}\label{defineGamma}
\hat{\Gamma}^0_{\lambda_0} = \{\gamma(t)\  \mbox{geodesics} \ : \ \dot\gamma^a \dot \gamma^b G_{ab} = 0, \ \mbox{and}\ l^a \dot \gamma^b G_{ab} = \lambda_0\}.
\end{equation}
In other words, the curves solving the equations of motions in the reduced system $(F, \omega, H(\lambda_0))$ have a one-to-one correspondence with null geodesics in $(\mathcal Q_{n+2}, G_{ab})$ satisfying $l_a \dot \gamma^a = \lambda_0$.  Reversely, the set of all null geodesics in  $(\mathcal Q_{n+2}, G_{ab})$ projects, by means of the map Eq.~\eqref{EisenhartProjection}, over the curves in $Q_n$ solving the equations of motion of the continuous family of reduced Hamiltonian systems $(F, \omega, H(\lambda))$, $\forall \ \lambda \in \mathbb R$.  So, there is a bridge between the dynamics of null geodesic in $(\mathcal Q_{n+2}, G_{ab})$ and the set of reduced Hamiltonian systems $(F, \omega, H(\lambda))$.

\subsection{Relation with Brinkmann spaces and pp-waves}

Until now we have done somethings that are well know for the physicist that works in the area of Classical Dynamics. However, thinking as a General Relativity physicist,  $(\mathcal Q_{n+2}, G_{ab})$ seems much more like some type of spacetime. Indeed, the  line element  Eq.~\eqref{BrinkmannMetric} appears in the framework of General Relativity and, also,  String Theory \cite{Tseytlin:1995fh}, as the most general metric we can build from a covariantly conserved null Killing vector field. In our case, this covariantly conserved vector is the one in Eq.~\eqref{PyFromGravity}: \begin{equation}\label{CovariantlyConservedVec}
 \nabla_a l_b = 0. 
\end{equation}
Spaces with a metric of the form Eq.~\eqref{BrinkmannMetric} are called \emph{Binkmann spaces} \cite{Brinkmann:1925fr}. Every $u = u_0$ hypersurface collapses to a  $n$-dimensional submanifold which is called the transverse space for $u = u_0$, and every transverse space corresponds to the configuration space $Q_n$ equipped with the transverse metric $\bar{g}_{ij}(q, u = u_0)$. \\

Following from the point of view of a physicist working in General Relativity, the next issue calling our curiosity is what about curvature. The first task is to compute the inverse metric $G^{ab}$. After some algebra, 
\begin{equation}\label{inverseEisenhartMet}
\begin{split}
 G^{uu} &= 0, \quad G^{uv} = 1 , \quad G^{ui} = 0 \\
 G^{vv} & = -  \Phi + \frac{1}{\bar g(n-1)!}\varepsilon^{i_1 \ldots i_n} \mathcal A_{i_1} \mbox{det}[\vec{\mathcal A},\vec{ g}_{i_2},\ldots, \vec{g}_{i_n}], \\
 G^{vi} & = -\frac{1}{\bar g(n-1)!} \varepsilon^{i j_2 \ldots j_n} \mbox{det}[\vec{\mathcal A}, \vec{g}_{j_2}, \ldots,\vec{g}_{j_n}], \\
 G^{ij} &=  \bar g^{ij},
\end{split}
\end{equation}
where $\vec{g}_k$ is the $k$-th row of $\bar g_{ij}$. Now, with some effort, the Christoffel symbols and, latter,  the Ricci tensor components can be computed. Fortunately, we can write them in terms of geometrical objects living at $Q_n$. Defining, 
\begin{equation} \label{DefVecTensor}
\begin{split}
C^i &=  \bar g^{ik}\left(\partial_u \mathcal A_k - \frac{1}{2}\partial_k \Phi\right), \quad
F_{ij} = \partial_i \mathcal A_j - \partial_j \mathcal A_i, \\
B_{ij} & = \frac{1}{2}(\partial_u \bar g_{ij} +  F_{ij}), 
\end{split}
\end{equation}
the non-vanishing components of the Ricci tensor are, 
\begin{equation}\label{RicciComponents}
\begin{split}
R_{uu} &= \nabla^{(\bar g)}_i C^i + \frac{1}{4} F^2 - \frac{1}{2}\left(\bar g^{ik}\partial_u^2 \bar g_{ik} + \frac{1}{2} \partial_u \bar g_{ik}\partial_u \bar g^{ik}\right), \\
R_{ui} &= \bar g^{jk}\nabla^{(\bar g)}_j B_{ik} - \frac{1}{2}\partial_i\left(\bar g^{jk}\partial_u \bar g_{jk}\right), \\
R_{ij} & = R_{ij}^{(\bar g)}, 
\end{split}
\end{equation}
where the superscript ${}^{(\bar g)}$ means "with respect to the metric $\bar g_{ij}$" for each value of $u$.  Finally, from Eq.~\eqref{inverseEisenhartMet}  and Eq.~\eqref{RicciComponents},
\begin{equation}\label{ScalarCurvature}
R = R^{(\hat g)}.
\end{equation}
So the scalar curvature is not affected by the scalar and vector potentials in the set of  reduced Hamiltonian systems $(F, \omega, H(\lambda))$, $\lambda \in \mathbb R$.\\

The results in Eqs.~\eqref{RicciComponents} and Eq.~\eqref{ScalarCurvature} do not  predict easy Einstein field equations, even for empty spacetime.  However,  for many many Hamiltonian systems, the kinetic term does not depends on the time $t$ and, even more, the kinetic metric $g_{ij}$ is flat. In this case, the Einstein field equations reduce to some like Riemannian Maxwell equations,
\begin{equation}\label{EinsteinFieldEquations}
\triangle^{(\bar g)}  \Phi = - \rho + \frac{1}{2} F^2 + 2 \partial_u \partial^k \mathcal A_k, \quad \nabla^{(\bar g) k} F_{ik} = J_i,
\end{equation}
where the energy density and current are defined as, 
\begin{equation}\label{currents}
\rho =  2 \kappa T_{uu}, \qquad J_i = 2\kappa T_{ui},
\end{equation}
where $\kappa = 8 \pi G_D$, and $G_D$ is the $D$-dimensional Newtonian gravitation constant. Additionally, $R= 0$ implies traceless stress tensor,
\begin{equation}\label{TracelessStressTensor}
T = G^{ab}T_{ab} = 0.
\end{equation}
The latter supposes a significant simplification from the general case. Indeed, in this case, the Brinkmann metric Eq.~\eqref{BrinkmannMetric} reduces to the one of a \emph{pp-wave}, revealing a connection between pp-waves and classical  Hamiltonian systems. Under this connection, the scalar potential $ \varphi = -\Phi/2$ gives the profile of the wave, while $A_i \ne 0$ is  related to the  presence of wave helicity\footnote{ Also known as gyratons, spacetimes describing the traveling of ultrarelativistic energy bunches with helicity.} \cite{Frolov:2005zq}. The energy density $\rho$ and current $J_i$ we need to generate the pp-wave can be read from the set of reduced Hamiltonian systems $(\mathcal F, \omega, H(\lambda))$ by means of the Eqs.~\eqref{EinsteinFieldEquations} and \eqref{currents}, so the spacetime and its matter content is entirely determined from the characteristics of the set of reduced classical mechanical system.

Finally with respect to Brinkmann spaces , they have a gauge symmetry with respect to shifts in the $v$ coordinate. The Killing vector $\vec{l}$ means that we have symmetry under translations $v\rightarrow v + v_0$. However, this symmetry can be gauged to, 
\begin{equation}\label{Gauge_sym_1}
v \rightarrow \bar v = v - \frac{1}{2} f(u, q). 
\end{equation}
Then, the line element Eq.~\eqref{BrinkmannMetric} remains unchanged provided that we transform $\Phi$ and $\mathcal A_i$ according to, 
\begin{equation}\label{Gauge_sym_2}
\Phi \rightarrow \bar \Phi =  \Phi + \partial_u f, \qquad  \mathcal A_i \rightarrow  \bar{\mathcal A}_i = \mathcal A_i + \partial_i f.
\end{equation}
In terms of the scalar potential $\varphi = -\Phi/2$ and the vector potential $A_i = \mathcal A_i/2$, these are the gauge transformations of electromagnetism.

\section{Eisenhart lift of autonomous and natural systems} \label{sec_3}

As a special example of the previous section, let us now consider the case in which  $(F, \omega, H(\lambda))$  is a set of  autonomous and natural systems, and $ g_{ij} = \delta_{ij}$ in standard Cartesian coordinates $x^i$. Then, $\mathcal Q_{n+2} = \mathcal Q_2 \times \mathbb R^n$, where $\mathcal Q_2$ is a $2$-dimensional manifold spanned by $u, v$ and $\mathbb R^n$ is the transverse space, while  the line element in Eq.~\eqref{BrinkmannMetric} reduces to, 
\begin{equation}\label{EisenhartForAutonomous}
ds^2 = 2 dv du +  \Phi(x) du^2 + d\vec{x}^2. 
\end{equation}
This line element can be also found for no-natural systems  by means of choosing a suitable null coordinate $\bar v$ in Eq.~\eqref{Gauge_sym_1}, such that the gauge transformations Eq.~\eqref{Gauge_sym_2}  can be used to vanishing $\mathcal A_i$ in Eq.~\eqref{BrinkmannMetric}, as long as $\mathcal A_i$ satisfies, 
\begin{equation}\label{Gauge_uses_1}
\partial_{[i}\partial_{j]} \mathcal A_k = 0, 
\end{equation}
and the line element Eq.~\eqref{EisenhartForAutonomous} is reachable provided that the transformed $\bar \Phi$  does not depend on $u$ and $R^{(g)} = 0$. In this situation, by Eqs.~\eqref{EinsteinFieldEquations} - \eqref{TracelessStressTensor} , the line element Eq.~\eqref{EisenhartForAutonomous} is sourced by a traceless stress tensor given by a energy density $\rho$ and vanishing current $J_i$, with field equation, 
\begin{equation}\label{Field_eq_autonomous}
\nabla^2 \Phi = -\rho.
\end{equation}
The independence of $\Phi$ with $u$ then implies $\rho = \rho(x^i)$. 
\\

Under the conditions previously discussed carrying to the line element Eq.~\eqref{BrinkmannMetric}, the Brinkmann space $(\mathcal Q_2 \times \mathbb R^n, G_{ab})$ has, at least, two Killing vectors: the vector field $\vec{l}$ in Eq.~\eqref{PyFromGravity} and $\vec{t} = \partial_u$. Since, 
\begin{equation}\label{ModuleT}
t^a t^b G_{ab} =  \Phi = -2 \varphi,
\end{equation}
the vector field $\vec{t}$ will be timelike everywhere provided that  $\varphi$ has a global positive minimum. Indeed, the existence of a global minimum of $\varphi$ is a desirable property we have to protect if we want to keep the quantum stability of the system, while the value of the minimum can be shifted to have a positive value without any consequence. In this way, always we can achieve to promote $(\mathcal Q_2 \times \mathbb R^n, G_{ab})$ to a stationary spacetime, where $\vec{t}$ can be used to define an observer $\vec{u}_p$ at each event $p \in \mathcal Q_2 \times \mathbb R^n$ as,  
\begin{equation}\label{ObserverField}
\vec{u}_p = \frac{1}{|\Phi|^{1/2}} \vec{t}. 
\end{equation}
In addition, requiring
\begin{equation}\label{FlatCond}
\lim_{x^i \rightarrow\infty} \Phi = -1 \quad \Rightarrow \quad \lim_{x^i \rightarrow\infty} \varphi = \frac{1}{2},  
\end{equation}
$(\mathcal Q_{2}\times \mathbb R^n, G_{ab})$ becomes asymptotically flat and, from Eq.~\eqref{ObserverField}, the vector field $\vec{t}$ defines the set of asymptotic observers. These are the observers measuring the charge and current defined in Eqs.~\eqref{currents}.\\

The Killing vector field $\vec{t}$ introduces a new constant of motion $t_a\dot \gamma^a$ that must be related to the conservation of energy in $(F, \omega, H(\lambda_0))$, since $u$ substitutes the time $t$ in Eq.~\eqref{LambdaSectorHamiltonian}. Indeed it is easy to compute that, 
\begin{equation}\label{ConservationPu}
 p_u = t^a \dot \gamma^b G_{ab}\,.
\end{equation}
Then, from Eq.~\eqref{LambdaSectorHamiltonian},
\begin{equation}\label{EnergyandPu}
t^a \dot \gamma^b G_{ab} = -\frac{E(\lambda_0)}{\lambda_0}, \quad \forall\ \gamma \in \hat \Gamma^0_{\lambda_0} (E), 
\end{equation}
where $\hat \Gamma^0_{\lambda_0} (E)\subset \hat \Gamma^0_{\lambda_0}$ is defined as the subset for trajectories with energy $E$ in the Hamiltonian system $(F, \omega, H(\lambda_0)$. The value of $p_u$ in Eq.~\eqref{ConservationPu} is always measurable for the set of asymptotic observers, so the energy of trajectories in each reduced Hamiltonian system $(F, \omega, H(\lambda))$ is reachable for the observers in the Brinkamnn space Eq.~\eqref{EisenhartForAutonomous}. \\

It is interesting to observe what happens with the geodesics of $\hat \Gamma^0_{\lambda_0}(E)$. In this case, the trajectories in the reduced system  are limited to a finite region of $\mathbb R^n$. From the point of view of the Brinkmann space, constant energy $E$ corresponds to the submanifold, 
\begin{equation}\label{constantEnergySurface}
\mathcal S_{E}(\lambda_0) = \left\{(u,v,x^i)\in \mathcal Q_2 \times \mathbb R^n :\ 2E + \lambda_0^2 \Phi(x) = 0\right\},
\end{equation}
such that the geodesics $\hat \Gamma^0_{\lambda_0}(E)$ are limited to the interior of $\mathcal S_E(\lambda_0)$. Indeed the normal vector to $\mathcal S_E(\lambda_0)$, $n_a = \partial_a(2E+\lambda_0^2  \Phi)$, satisfies, 
\begin{equation}\label{NormalVectorTangent}
n_a \dot \gamma^a = \left.\lambda_0^2 \dot x^i \partial_i  \Phi\right|_{\mathcal S_E(\lambda_0)}= 0, \quad \forall\ \gamma \in \hat \Gamma^0_{\lambda_0}(E). 
\end{equation}
So the geodesics of $\hat \Gamma_{\lambda_0}^0(E)$ reach $\mathcal S_E(\lambda_0)$ tangentially. Also note that,
\begin{equation}\label{NormalVectorNorma}
n_a n^a = \bar g^{ij}\partial_i  \Phi \partial_j  \Phi > 0, 
\end{equation}
since $\bar g_{ij}$ is a Riemannian metric. Thus, $\mathcal S_E(\lambda_0)$ is a time-like submanifold everywhere.\\


\begin{figure*}[!t]
\centering
\begin{tabular}{ccc}
\includegraphics[width=0.45\textwidth, height=0.30\textheight]{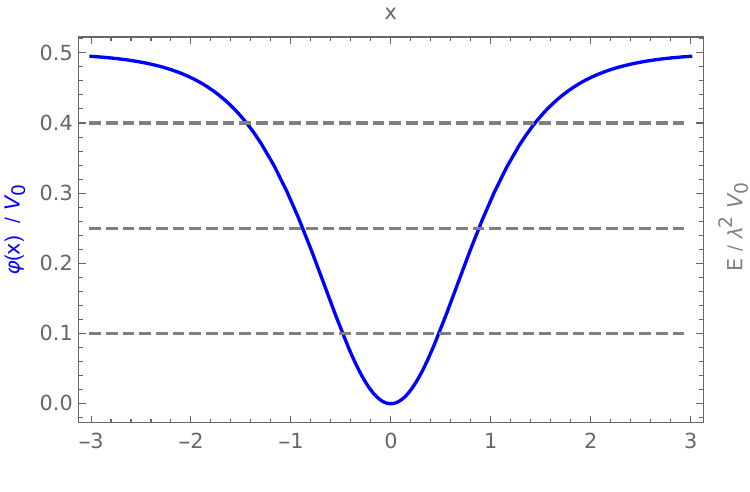}
&
\hspace*{0.2cm}
&
\includegraphics[width=0.45\textwidth, height=0.30\textheight]{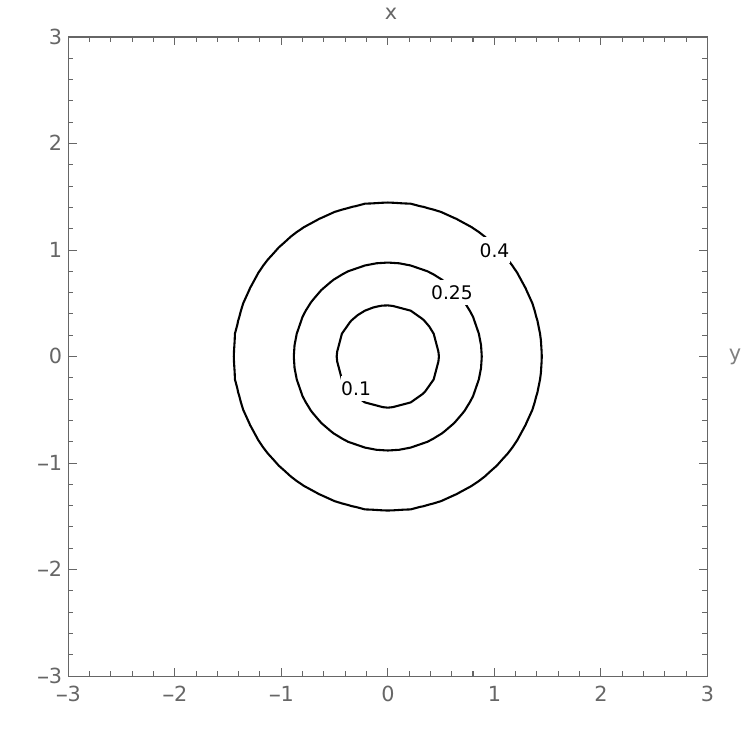}
\end{tabular}
\caption{\label{fig1} \emph{Left Panel.-} Pöschl-Teller potential Eq.~\eqref{AsymptoticWell} for $k =1$. \emph{Right Panel.-} Projection of surfaces $\mathcal S_E(\lambda)$ for $E/\lambda^2 V_0 = 0.1, 0,25, 0.4$ in the plane $XY$.}
\end{figure*}


As an example, we can consider the isotropic harmonic oscillator potential, 
\begin{equation}\label{HarmonicPotential}
\varphi_{_{HO}}(x^i) = \frac{\vec{x}^2}{2}.
\end{equation}
The potential has a global minimum at $x^i = 0$, with $\varphi_{{}_{HO}}^{(min)} = 0$, and $\mathcal S_E(\lambda)$ are time-like hypersurfaces wrapping around $\mathcal Q_2 \times \{\vec{0}\}$ which project as concentric $(n-1)$-spheres in the transverse space $\mathbb R^n$. The value of $\lambda^{-1}$ scales the radius of the hypersurfaces $\mathcal S_E(\lambda)$ and, for the same energy $E$, geodesics with lower value of $l^a \dot \gamma^b G_{ab}$ reaches further points in $\mathbb R^n$. However, the potential violates Eq.~\eqref{FlatCond} and $(\mathcal Q_2 \times \mathbb R^n, G_{ab})$ fails to be asymptotically flat, without a set of asymptotic observers. The situation can be corrected by considering the Pöschl-Teller potential,
\begin{equation}\label{AsymptoticWell}
\varphi_{{}_{PT}}(x^i) =  - V_0\left[ \frac{1}{2 \cosh^2(k|\vec{x}|)} - \frac{1}{2}\right],\quad k>0. 
\end{equation}
This potential has a global minimum at $x^i = 0$, with $\varphi_{{}_{PT}}^{(min)} = 0$, and also has a finite asymptotic behaviour, 
\begin{equation}\label{Asymptotic_Behaviour_VPT}
\lim_{x^i\rightarrow \infty } \varphi_{{}_{PT}} (\vec{x}) = \frac{V_0}{2}. 
\end{equation}
Then, redefining coordinates $u$ and $v$ as,
\begin{equation}\label{redefineUandVforCorrectDimensions}
\sqrt{V_0}\ u \rightarrow u, \quad \frac{v}{\sqrt{V_0}} \rightarrow v, 
\end{equation}
also satisfies the asymptotic flatness condition Eq.~\eqref{FlatCond}. Therefore, for fixed $\lambda_0$ and energies $0 < E < \lambda_0^2 V_0/2$, the  projection of the submanifolds $\mathcal S_E(\lambda_0)$  over $\mathbb R^n$ are concentric $(n-1)$-spheres, while for energies $E > \lambda_0^2 V_0/2$ the null geodesics of $\hat\Gamma^{0}_{\lambda_0} (E)$ escape to infinite (see Fig~\ref{fig1}).

\section{Quantum dynamics from Eisenhart lift} \label{sec_4}

In the previous sections we have shown that the projection of the sets of geodesics $\hat \Gamma^0_{\lambda}$, $\lambda \in \mathbb R$, over the transverse space of the Brinkmann space $(\mathcal Q_{n+2}, G_{ab})$ corresponds to the curves solving the equations of motion of the set of reduced Hamiltonian systems $(F, \omega, H(\lambda))$, with the time $t$ being the affine parameter of the curves in $\hat \Gamma^0_\lambda$. In this section we propose to define quantum states for null particles in $(\mathcal Q_{n+2}, G_{ab})$ from Quantum Mechanics in the transverse space, \emph{i.e.} proposing that the space of quatum states for null geodesics in $(\mathcal Q_{n+2}, G_{ab})$ projects over the Hilbert space of states in the transverse space $( Q_n, \bar g_{ij})$ for each value of $\lambda$, such that the continuous set of reduced Hamiltonian systems $(F, \omega, H(\lambda))$ governs the time evolution of quatum null geodesics in $(\mathcal Q_{n+2}, G_{ab})$.\\

Time evolution must be related to a set of well defined observers. Thus, for simplicity, from now on we shall constrain our analysis to Brinkmann spaces  with a transverse metric that  does not depend on $u$ and, also, such that $\vec{t} = \partial_u$  describes a set of well defined asymptotic observers. Under these conditions, the transverse space  $( Q_n, \bar{g}_{ij})$ corresponds to 'time slices' -which do not change with the timelike coordinate $u$-  where Quantum Mechanics is well posed and states can be built by means of measures performed by the set of asymptotic observers. Then, in a coordinate representation, we define operators $\hat p_i$ as,  
\begin{equation}\label{operatorsPandQ}
\hat q^i = q^i,\quad  \hat p_i = -i\left[\partial_i + \sigma_i(q)\right] 
\end{equation}
where, following Ref.~\cite{RevModPhys.29.377}, $\sigma_i(q)$ are the coefficients of the connection on the bundle over $Q_n$ with fiber $\mathbb C$, such that wave functions are section of this fiber bundle. They are given by: 
\begin{equation}\label{YangMillsFields}
\mathfrak{Re}(\sigma_i) = \frac{1}{4} \hat g^{jk}\partial_i \hat g_{jk}.
\end{equation}
The coefficients  $\sigma_i(q)$ are introduced to guarantee that $\hat p_i$ are self-adjoint and the commutation rules, 
\begin{equation}\label{conmutationrelation}
[\hat q^i,\hat q^j ] = 0, \quad [\hat p_i,\hat q^j ] = i\ \delta_i^j, \qquad [\hat p_i, \hat p_j] = 0. 
\end{equation}
are satisfied\footnote{See Ref.~\cite{kleinert2009path} for a modern introduction to the subject of Quantum Mechanics in Riemannian manifolds}. In addition, in order to lift the quantum dynamics from the transverse space to the Brinkmann space we have to introduce also an operator $\hat p_v$ such that, 
\begin{equation}\label{KetRepresentationPv}
\hat{p}_v \ket{\Psi_{\lambda_0}}
= \lambda_0 \ket{\Psi_{\lambda_0}},
\end{equation}
where $\ket{\Psi_{\lambda_0}}$ is the state of a null particle following a geodesic with $p_v = \lambda_0$. In a coordinate representation, 
\begin{equation}\label{CoordinateRepresentationPv}
\hat{p}_v = -i \partial_v\quad \Rightarrow \quad \Psi_{\lambda_0}(v,q) = \frac{1}{\sqrt{2\pi}}e^{i\lambda_0 v} \varphi(q), 
\end{equation}
such that $[\hat p_v, \hat v] = i$ is satisfied. Here $\varphi(q)$ is a wave function over the transverse space $(Q_n, \bar{g}_{ij})$ that, under time evolution in the Schrödinger picture, obeys the Hamiltonian $H(\lambda_0)$.  \\

The states $\ket{\Psi_{\lambda_0}}$ span a Hilbert space, 
\begin{equation}\label{HilbertSpaceForLambda0}
{\mathfrak{H}}_{\lambda_0} = \ket{\lambda_0}\otimes \mathfrak H, 
\end{equation}
where $\mathfrak{H}$ is the Hilbert space of states over the transverse space $(Q_n, \bar{g}_{ij})$ and $\ket{\lambda_0}$ is the state living over the line spanned by $v$ with $\hat p_v \ket{\lambda_0} = \lambda_0 \ket{\lambda_0}$ , 
\begin{equation}\label{EigenState_lambda_0}
\braket{v|\lambda_0} = e^{i\lambda_0 v},
\end{equation}
where $\ket{v}$ means "localized at $v$". Now we build the bundle $(\mathcal B, \bar \pi, \mathbb R)$, where, 
\begin{equation}\label{HilbertBundel}
\mathcal B =  \bigcup_{\lambda \in \mathbb R}{\mathfrak{H}}_{\lambda} \simeq \mathbb R \times \mathfrak{H},
\end{equation}
such that $\bar{\pi}^{-1} (\lambda_0) = \mathfrak{H}_{\lambda_0}$. Therefore, each fiber of the bundle Eq.~\eqref{HilbertBundel} is isomorphic to the Hilbert space $\mathfrak H$ but, with respect to time evolution, states in the fiber at $\lambda_0$ are related to the Hamiltonian $H(\lambda_0)$. Then, finally we define null particle states in a Brinkmann space as averaged sections in $(\mathcal B, \bar{\pi}, \mathbb R)$ over $\mathbb R$, such that the projection of an state $\ket{\Psi}$ over the transverse space is given by,  
\begin{equation}\label{states}
\braket{v|\Psi} = \frac{1}{\sqrt{2\pi}} \sum_j\int_{\mathbb R} d\lambda\ C_j(\lambda) e^{i\lambda v} \ket{\varphi_j}_\lambda,
\end{equation}
where  $\ket{\varphi_j}_\lambda$ is a discrete basis of $\mathfrak{H}$ subjected to time evolution by means of the Hamiltonian $H(\lambda)$. In this way, $\ket{\Psi}$ lives at the space, 
\begin{equation}\label{Direct_Sum_Hilbert}
\ket{\Psi} \in \bigoplus_{\lambda \in \mathbb R} \mathfrak H_\lambda.
\end{equation}
The Hilbert product extends naturally to the kets $\ket{\Psi}$ as, 
\begin{equation}\label{HilbertInBrinkmann}
\braket{\Psi_1|\Psi_2} = \int_{-\infty}^{\infty} dv\, \braket{\Psi_1|v}\braket{v|\Psi_2} =  \sum_{j,k}\int_{\mathbb R} d\lambda \ C^{(1)*}_j(\lambda) C_k^{(2)}(\lambda) \braket{\varphi_j|\varphi_k}_\lambda,
\end{equation}
In particular, we recover the wave function Eq.~\eqref{CoordinateRepresentationPv} for $C_j(\lambda) =  \delta(\lambda - \lambda_0)$ for some $j$. Respect to normalization of kets,  
\begin{equation}\label{Norm_Psi}
\braket{\Psi |\Psi} = \sum_{j,k}\int_{\mathbb R} d\lambda \ C^{*}_j(\lambda) C_k(\lambda) \braket{\varphi_j|\varphi_k}_\lambda,
\end{equation}
which fixes, 
\begin{equation}\label{Norm_condition_Psi}
\sum_j \int_{\mathbb R} d\lambda \ |C_j(\lambda)|^2 = 1, 
\end{equation}
if the basis $\{\ket{\varphi_j}\}$ of each $\mathfrak H_\lambda$ is orthonormal. Finally, given some observable $\hat O(v)$, 
\begin{equation}\label{Expect_value}
\begin{split}
&\braket{\hat O}_{\Psi} =  \\
&\frac{1}{2\pi}\sum_{j,k}\int_{-\infty}^{\infty} dv \int_{\mathbb R^2} d\lambda_1 d\lambda_2 \ C^{(1)*}_j(\lambda_1) C_k^{(2)}(\lambda_2) e^{i(\lambda_2 - \lambda_1)v} \braket{\varphi_j|\hat O(v)|\varphi_k}_\lambda.
\end{split}
\end{equation}
Note that, for $\hat O(v)$ changing slowly with $v$ with characteristic length $\Delta v$, this expected value can be approximated by, 
\begin{equation}\label{Slow_Oscillating_approx}
\braket{\hat O}_{\Psi} = \frac{1}{\Delta v}\sum_{j,k}\int_{-\frac{\Delta v}{2}}^{\frac{\Delta v}{2}} dv \int_{\mathbb R} d\lambda \ C^{(1)*}_j(\lambda) C_k^{(2)}(\lambda)  \braket{\varphi_j|\hat O(v)|\varphi_k}_\lambda+ F.O.T.
\end{equation}
where $F.O.T$ means fast oscillating terms. \\

Together with pure states Eq.~\eqref{states}, mixed states can be also considered. For a mixed state, the density operator will be,  
\begin{equation}\label{densityOperatorDef}
\bar \rho = \sum_j \int_{\mathbb R} d\lambda \  w_j (\lambda) \ket{\Psi_j}_\lambda\bra{\Psi_j}_\lambda,
\end{equation}
where, 
\begin{equation}\label{DensityOppCoef}
\braket{v|\Psi_j}_\lambda = e^{i\lambda v} \ket{\varphi_j}_\lambda, \qquad \sum_j \int_{\mathbb R}  d\lambda\ w_j(\lambda) = 1. 
\end{equation}
Thus, 
\begin{equation}\label{DensityOpp}
\bar \rho = \sum_j \int_{\mathbb R} d\lambda\ w_j(\lambda) \ket{\varphi}_\lambda \bra{\varphi}_\lambda
\end{equation}
For the special case that $w_j(\lambda)$ factorizes as, 
\begin{equation}\label{SpecialFactorization}
w_j(\lambda) = p_j\ w(\lambda), \quad \sum_j p_j = 1, 
\end{equation}
The density operators take the more simple form, 
\begin{equation}\label{specialCaseDensityOpp}
\bar{\rho} = \int_{\mathbb R} d\lambda\   w(\lambda) \rho_{\lambda}, \quad w(\lambda) \in [0,\infty) \ : \ \int_{\mathbb R} d\lambda\ w(\lambda) = 1, 
\end{equation}
where, 
\begin{equation}\label{DenstOppForEachLambda}
\rho_\lambda = \sum_j p_j \ket{\varphi}_\lambda \bra{\varphi}_\lambda,
\end{equation}
is a density matrix on  the Hilbert space $\mathfrak H$ evolving with $H(\lambda)$, such  that the state Eq.~\eqref{specialCaseDensityOpp} projects over density matrices $\rho_\lambda$ with the same probabilities $\{p_j\}$ for each value of $\lambda$.\\

\subsection{Massless particle states in stationary Brinkmann spacetimes} 

In general terms, we cannot add much more to the quantumness of null geodesics in Brinkmann spaces only subjected to the constrains of well defined asymptotic observers and the no $u$-dependence of the transverse metric $\bar g_{ij}$. However, by adding the condition that $(\mathcal Q_{n+2},G_{ab})$ be stationary with $\vec{t} = \partial_u$ a time-like Killing, we can also discuss ground states, stationary states and time evolution since, under these conditions, the set $\{H(\lambda)\}$  is a family of  autonomous Hamiltonians. \\

Naively we could think that the ground state for the QM dynamics of null particles we are constructing can be defined from the ground state in each fiber of Eq.~\eqref{HilbertBundel}.  However, each fiber $\mathfrak H_\lambda$ has a ground state $\ket{0}_\lambda$ and, thus, there is no way to define uniquely a ground state for the Hilbert space \eqref{Direct_Sum_Hilbert}. That makes sense since time evolution in \eqref{Direct_Sum_Hilbert} does not come from one unique Hamiltonian, but from the complete family of reduced Hamiltonians $\{H(\lambda),\ \lambda \in \mathbb R\}$. For example, for any fixed $\lambda_0$, the state given by, 
\begin{equation}\label{Ground_define_lambda}
\braket{v|\Omega}_{\lambda_0} = \frac{1}{\sqrt{2\pi}} e^{iv\lambda_0} \ket{0}_{\lambda_0}, 
\end{equation}
is a ground state, with definite momentum $p_v = \lambda_0$, in the sense that there is no possibility to connect it with a lower energy state by means of operators in our model.   In general, any ground state of the model will respond to an expression of the form:    
\begin{equation}\label{General_Ground_State}
\braket{v|\Omega} = \frac{1}{\sqrt{2\pi}} \int_{\mathbb R} d\lambda C_0(\lambda) e^{i\lambda v} \ket{0}_\lambda,
\end{equation}
such that, 
\begin{equation}\label{Norm_ground}
\int_{\mathbb R}d\lambda\, \left|C_0(\lambda)\right|^2 = 1.
\end{equation}
As an example, the expect value of $\hat p_v$ over Eq.~\eqref{General_Ground_State}, 
\begin{equation}\label{VEV_p_v}
\braket{\hat p_v}_\Omega = \int_{\mathbb R} d\lambda |C_0(\lambda)|^2 \lambda,
\end{equation}
And , therefore, it is possible to build ground states with zero expected value of $\hat p_v$ providing $|C_0(\lambda)|$ be a defined parity function.\\

Together with the fact that, in our model,  there is an infinite of ground states for the quantum dynamics of null particles in Brinkmann spaces, the existence of a continuous family of Hamiltonians $\{H(\lambda)\}$ over the transverse space $(Q_{n}, \bar g_{ij})$ also leads to different time evolution in the Schrödinger picture for each $\lambda$-component in Eq.~\eqref{states}. For each state $\ket{\Psi_{\lambda_0}}$ of definite momentum $p_v = \lambda_0$, 
\begin{equation}\label{Time_evolution1}
i \partial_t \braket{v|\Psi_{\lambda_0}} = \frac{1}{\sqrt{2\pi}} e^{i\lambda_0 v} \hat H(\lambda_0) \ket{\varphi},
\end{equation}
where $t$ is the affine parameter over null geodesics satisfying the Eqs.~\eqref{equationsOfMotion} and \eqref{CompleteEqsOfMotion}. Now, since $t$ runs with the coordinate $u$ as $\dot u = \lambda_0$, 
\begin{equation}\label{Time_evolution2}
i \partial_u \braket{v|\Psi}_{\lambda_0} = \frac{1}{\sqrt{2\pi}} \frac{e^{i\lambda_0 v}}{\lambda_0} \hat H(\lambda_0) \ket{\varphi}.
\end{equation}
For the case that  $\ket{\varphi}$ be an eigenvector of $H(\lambda_0)$ with energy $E_0$, 
\begin{equation}\label{Time_evolution3}
i \partial_u \braket{v|\Psi} _{\lambda_0}= \frac{1}{\sqrt{2\pi}} \frac{e^{i\lambda_0 v}}{\lambda_0} E_0 \ket{\varphi} = \frac{E_0}{\lambda_0} \braket{v|\Psi}_{\lambda_0}, 
\end{equation}
and  $\braket{v|\Psi_{\lambda_0}}$ describe a  stationary state. As a particular example, we can consider the ground state Eq.~\eqref{Ground_define_lambda}.  Since near a minimum of $\varphi(q^i)$, we can approximate the potential by  the harmonic oscillator one, 
\begin{equation}\label{Time_evolution4}
i \partial_u \braket{v|\Omega}_{\lambda_0} = \frac{\lambda_0 \omega_0}{2 }\braket{v|\Omega}_{\lambda_0}, \qquad  \omega_0^2  = \sum_{i=1}^n h_i,
\end{equation}
where $h_i$ are the eigenvalues of the Hessian of $\varphi(q_i)/2$ in the minimum. In general, for a pure state responding to the expression Eq.~\eqref{states}, 
\begin{equation}\label{Time_evolution6}
i \partial_u \braket{v|\Psi} = \frac{1}{\sqrt{2\pi}}  \sum_j\int_{\mathbb R} d\lambda \frac{C_j(\lambda)}{\lambda} e^{i\lambda v} \hat H(\lambda) \ket{\varphi_j}_\lambda. 
\end{equation}
Therefore, for $\ket{\varphi_j}_\lambda$ eigenvectors of $\hat H(\lambda)$ satisfying,
\begin{equation}\label{Time_evolution7}
\hat H(\lambda) \ket{\varphi_j}_\lambda \sim \lambda \ket{\varphi_j}_\lambda,\quad \forall\ \lambda \in \mathbb R
\end{equation}
the pure state Eq.\eqref{states} is stationary.

\section{Quantum mechanical phenomena in stationary Brinkmann spacetimes} \label{sec_5}

If the quantum dynamics of null particles in Brinkmann spaces works according to the model developed in the previous section, quantum phenomena would appear for low enough energies. For the case of the Brinkmann space given by Eq.~\eqref{EisenhartForAutonomous}, the only nonvanishing component of the Ricci tensor Eq.~\eqref{RicciComponents}, $R_{uu}$, scales as second derivatives of $\Phi$. Thus, 
\begin{equation}\label{Quantum_phenomena_1}
T_{uu} \sim \Phi''/2\kappa.
\end{equation}
On the other hand, near a minimum $q_0$ of $V(q)$ (therefore a maximum of $\Phi$), the energy in the transverse space of the null geodesic with $p_v = \lambda$ scales as, 
\begin{equation}\label{Quantum_phenomena_2}
E^2 \sim \lambda^2 \Phi''.
\end{equation}
Therefore, in Planck units, the scale of the energy density to observe quantum phenomena at an energy scale $E$ in the transverse space is, 
\begin{equation}\label{Quantum_phenomena_3}
\frac{T_{uu}}{\rho_p} \sim \frac{1}{\lambda^2}\left(\frac{E}{E_p}\right)^2,
\end{equation}
where $\rho_p \sim 10^{96}$ (SI) is the Planck density and $E_p \sim 10^9$ (SI) the Planck energy. In terms of the energy density defined in Eq.~\eqref{currents},
\begin{equation}\label{Quatum_phenomena_4}
\rho \sim \rho_p\  \frac{\kappa}{\lambda^2} \left(\frac{E}{E_p}\right)^2,
\end{equation}
or, alternatively, from the Ricci tensor in Eq.~\eqref{RicciComponents}, 
\begin{equation}\label{Quatum_phenomena_5}
R_{uu} \sim \rho_p \frac{\kappa}{\lambda^2} \left(\frac{E}{E_p}\right)^2.
\end{equation}
So, for large enough $\lambda$, quantum phenomena for null particles  appears in a regime where the gravity can still be classical.\\   

With respect to quantum phenomena, there are two characteristic phenomena of quantum mechanics we can consider: quantum tunneling and entanglement.  

\subsection{Quantum  tunneling}

\begin{figure*}[!t]
\centering
\begin{tabular}{ccc}
\includegraphics[width=0.45\textwidth, height=0.30\textheight]{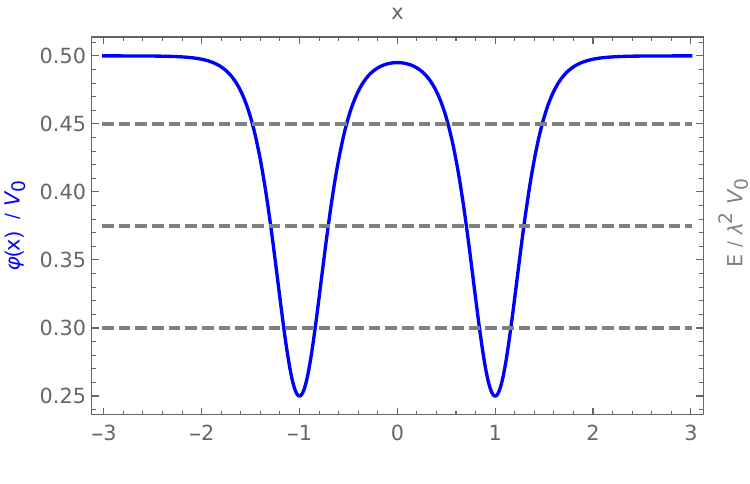}
&
\hspace*{0.2cm}
&
\includegraphics[width=0.45\textwidth, height=0.30\textheight]{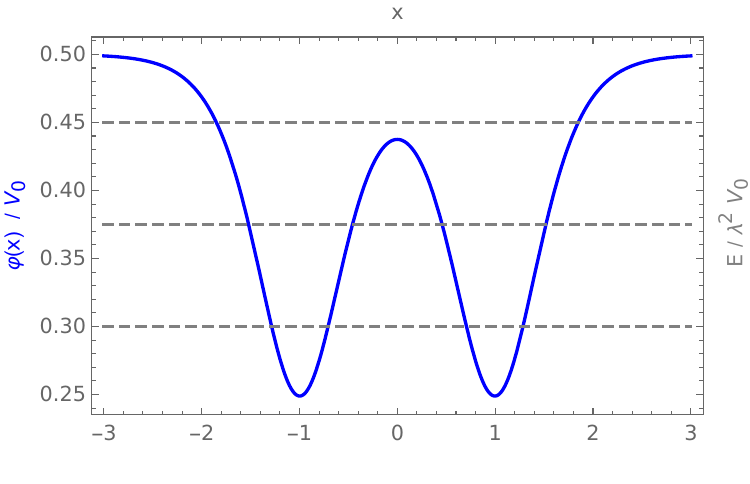}
\\
\includegraphics[width=0.45\textwidth, height=0.30\textheight]{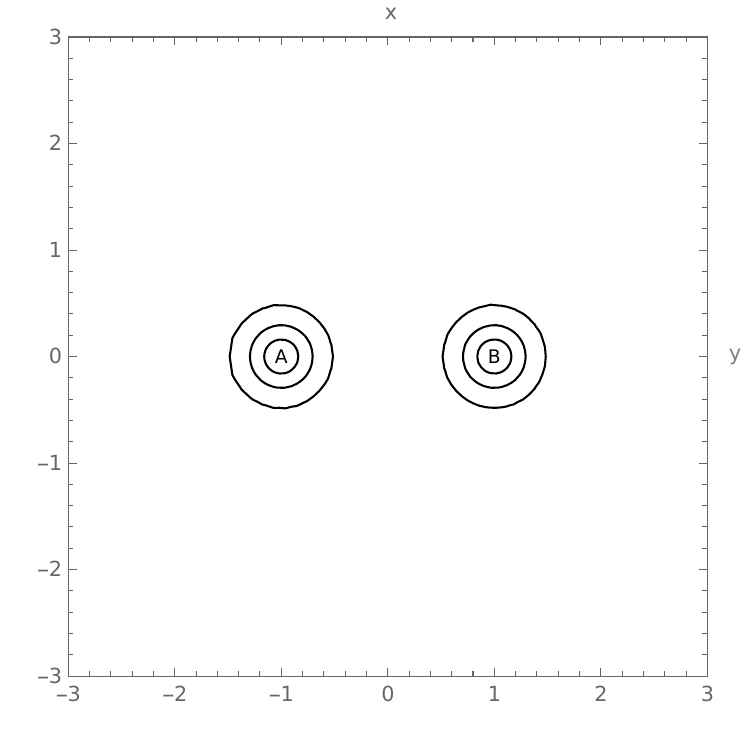}
&
\hspace*{0.2cm}
&
\includegraphics[width=0.45\textwidth, height=0.30\textheight]{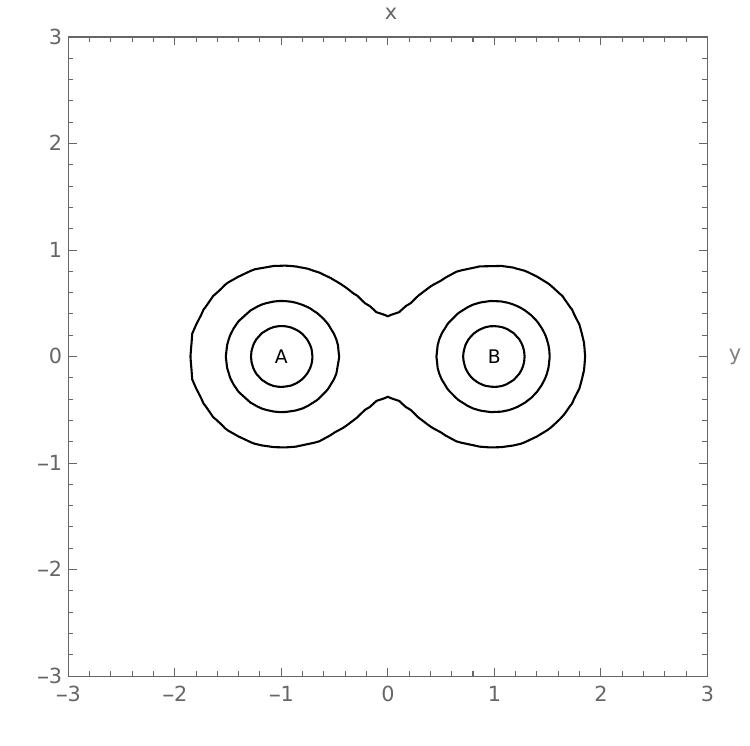}
\end{tabular}
\caption{\label{fig2} Double well potential Eq.~\eqref{DoublePoschlTeller}  (up) and projection of surfaces $\mathcal S_E$ (down) for $E/\lambda^2 V_0 = 0.45, 0.375, 0.3$ in the plane $XY$ for $\vec{x}_0 = (1,0)$. \emph{Left column ($k = 3$)}.- Observers in regions $A$ and $B$ cannot communicate by means of geodesics in $\hat\Gamma_\lambda^0(E)$ for any of the three proposed values of $E$. Surfaces $\mathcal S_E$ split in two pieces for any of the three values of energy. \emph{Right column ($k = 1.5$)}.- Observers in region $A$ and $B$ can communicate by means of geodesics in $\hat\Gamma^0_\lambda(E)$ for $E/\lambda^2 V_0= 0.3$, but they cannot for $E/\lambda^2 V_0 = 0.375, 0.45$. As a consequence, regions $A$ and $B$ become together for $E/\lambda^2 V_0 = 0.3$.}
\end{figure*}

In Sec.~\ref{sec_3} we showed the example of a stationary Brinkmann space, with a well defined set of asymptotic observers, where null geodesics of $\hat \Gamma^0_{\lambda_0} (E)$ were trapped inside the region bounded by the hypersurface $\mathcal S_E(\lambda_0)$ (see Fig.~\ref{fig1}). For that we propose the Pöschl-Teller potential Eq.~\eqref{AsymptoticWell} for the set of reduced Hamiltonian systems $(Q_n, \omega, H(\lambda))$. A more interesting situation arises for the case the potential $\varphi$ has several minima. In this case, $\mathcal S_E(\lambda_0)$ could be a disconnected submanifold composed of two or more hypersurfaces deppending on the energy $E$. As an example, consider  the double well potential, 
\begin{equation}\label{DoublePoschlTeller}
 \varphi_{{}_{2PT}}(x^i) = \frac{1}{2}\left[ \varphi_{{}_{PT}}(x^i - x_0^i) + \varphi_{{}_{PT}}(x^i + x_0^i) \right] , 
\end{equation}
This potential satisfies the condition Eq.~\eqref{FlatCond} and, for $k |\vec{x_0}| >>1$, has two minima located at, 
\begin{equation}\label{MinimaDoublePoschteller}
x^i_ \pm \simeq \pm x^i_0, 
\end{equation}
such that, 
\begin{equation}\label{VMinima}
 \varphi_{{}_{2PT}}(x^i_\pm) = \frac{V_0}{2} \left[\frac{1}{2} -\frac{1}{2\cosh (2k|\vec{x_0}|)}\right] \simeq \frac{V_0}{4}.
\end{equation}
Also, the potential has a maximum located between the to minima, at $x= 0$, with value, 
\begin{equation}\label{MaximaDoublePot}
\varphi_{{}_{2PT}}^{(max)} \simeq V_0/2,
\end{equation}
for $k |\vec{x_0}| >>1$. Then, for fixed $\lambda_0$ and energies satisfying $\lambda_0^2 V_0/4 <E<\lambda_0^2 V_0/2$,  $\mathcal S_{E} (\lambda_0)$ could splits in two disconnected hypersurfaces, as it is shown in Fig.~\ref{fig2}, which enclose separate regions, $A$ and $B$ of the Brinkmann space $(\mathcal Q_2\times R^n, G_{ab})$.  For two observers Eq.~\eqref{ObserverField} located at events $p_1 \in A$ and $p_2\in B$ that pretend to communicate through light signals, they cannot do if they use the "$\lambda_0$-channel" (geodesics satisfying $p_v = \lambda_0$) for energies below some threshold $E_0 = \varphi_{{}_{2PT}}^{(max)}$. \\


The classical behaviour we have exposed changes in the framework of the quantum mechanical model developed in Sec.~\ref{sec_4} since, then, quantum tunneling between regions $A$ and $B$ appears. Let us consider the case of  a generic double-well potential $\varphi(x)$ over a one-dimensional transverse space $\simeq \mathbb R$, and such that $\varphi(x)$ possesses two symmetric minima at $\pm x_0$. Then, the tunneling between the two minima for null particles with $p_v = \lambda$, in the semiclassical approximation,  can either be approached  from the paradigm of instanton physics or the WKB method \cite{Coleman:1978ae, PhysRevD.16.423, SONG20082991}. Choosing the first option, the penetration factor of the classically forbidden region in a (Euclidean) time $T$ is given by, 
\begin{equation}\label{Tunneling_1}
 \Delta = e^{- \lambda S_0 T}, 
 \end{equation}
where $S_0$ is the one-instanton action minus the action of the solution corresponding to the equilibrium ($x = \pm x_0$)\footnote{Usually instantons in the double-well potential problem are computed for potentials satisfying $\varphi(\pm x_0) = 0$ and, thus, $S[x=\pm x_0] = 0$. That is not our case.}. In terms of the $uu$-component of the metric Eq.~\eqref{EisenhartForAutonomous}, $\Phi (x)$, it is: 
\begin{equation}\label{Tunneling_2}
S_0 = \int_{-x_0}^{x_0} dx\  \left[\Phi (x_0) - \Phi(x) \right]^{1/2}, 
\end{equation}
Then, it is a known result that under the dilute instanton gas condition and for large $T$ ,  
\begin{equation}\label{Tunneling_3}
\begin{split}
\braket{\pm x_0| e^{-H(\lambda) T}| -x_0 } &\simeq \frac{1}{2}\left(\frac{\lambda \omega}{\pi }\right)^{1/2} e^{- \lambda \omega T/2} \\
&\times\left[\mbox{exp}\left(K T e^{- \lambda S_0 }\right)\mp \mbox{exp}\left(-K T e^{- \lambda S_0} \right)\right],
\end{split}
\end{equation}
where $\omega$ is given by, 
\begin{equation}\label{Tunneling_4}
\omega^2 = -\frac{1}{2} \Phi''(x_0), 
\end{equation}
and the constant factor $K$ is computed from the one-instanton solution, 
\begin{equation}\label{Tuneling_5a}
t = t_0 + \frac{1}{\lambda }\int_0^{\bar x(t)} dx\  \left[\Phi (x_0) - \Phi\left(x\right) \right]^{-1/2}
\end{equation}
as, 
\begin{equation}\label{Tuneling_5}
K = \left(\frac{\lambda S_0}{2 \pi}\right)^{1/2}  \left|\frac{\mbox{det} \left(- \partial^2_t + \lambda^2 \omega^2\right)}{\mbox{det'}\left(- \partial^2_t - \frac{\lambda^2}{2}\Phi''(\bar x)\right)}\right|.
\end{equation}
where the prime determinant means it may be computed excluding the zero eigenvalue\footnote{See \cite{Coleman:1978ae} for more details.}. 

From Eq.~\eqref{Tunneling_3} it is also possible to read the ground state energy for each $\lambda$ \cite{PhysRevD.16.423}. Naively, we could think that we have a double degenerate ground state for each $\lambda$, since $V_{2PT}$ has two symmetric minima. Let us to call $\ket{0}_R$ and $\ket{0}_L$  to the ground states corresponding to each minima.  However, because tunneling, this expected degeneracy breaks. By the symmetry $\pm x_0 \leftrightarrow \mp x_0$, $\ket{0}_R$ and $\ket{0}_L$  combine into odd and even states, 
\begin{equation}\label{Tunneling_6}
\ket{-} = \frac{1}{\sqrt{2}}\left(\ket{0}_R - \ket{0}_L\right), \qquad \ket{+} = \frac{1}{\sqrt{2}}\left(\ket{0}_R + \ket{0}_L\right), 
\end{equation}
that correspond to the ground and first excited states, with energies $E_\pm$. On the other hand, in a (discrete) basis of $H(\lambda)$, 
\begin{equation}\label{Tunneling_7}
\braket{+x_0 | e^{-H(\lambda) T}| - x_0} = \sum_n e^{-E_n(\lambda) T}\braket{x_0|n}\braket{n|x_0},
\end{equation}
Thus, the leading order term and the first next-to-leading-order term in Eq.~\eqref{Tunneling_7} are, 
\begin{equation}\label{Tunneling_8}
\braket{+x_0 | e^{-H(\lambda) T}| - x_0} \simeq e^{-E_- T} \Psi^*_-(-x_0)\Psi_-(x_0) + e^{-E_+ T} \Psi^*_+(-x_0)\Psi_+(x_0). 
\end{equation}
Then, by comparing with Eq.~\eqref{Tunneling_3}, in the semi-classical limit the ground state energy and first excited energy are, 
\begin{equation}\label{Tunneling_9}
E_\pm(\lambda) \simeq  \frac{\lambda\omega}{2} \pm KT e^{-\lambda S_0}.
\end{equation}
Thus, the energy split because of tunneling between the ground state and the first excited state for each $\lambda$ is proportional to the penetration factor in the classically forbidden  region.

As an example, we can compute the tunneling splitting for the special case of the double-well potential given at Eq.~\eqref{DoublePoschlTeller}. The minima of the potential are given by the zeroes of: 
\begin{equation}\label{Tuneling_esp_1}
\Phi'(x) = -2 k V_0 \left\{\frac{\sinh\left[2k(x-x_0)\right]}{\left(\cosh \left[2k(x-x_0)\right] + 1\right)^2} +\frac{ \sinh\left[2k(x+x_0)\right]}{\left(\cosh \left[2k(x+x_0)\right] + 1\right)^2}\right\}.
\end{equation}
For large $kx_0>> 1$, at $x=\pm x_0$ we have, 
\begin{equation}\label{Tuneling_esp_2}
\Phi'(\pm x_0) \simeq - 2 k V_0 \frac{\tanh(\pm 4 k x_0)}{\cosh(4kx_0)} = 0 + \mathcal O\left(1/\cosh(4kx_0)\right).
\end{equation}
So, we can take $x= \pm x_0$ as the minima until order $1/\cosh(4kx_0)$. Then, 
\begin{equation}\label{Tuneling_esp_3}
\Phi''(\pm x_0) = - \simeq  k^2 V_0 \left[\frac{1}{2} + \frac{4}{\cosh(4kx_0)}\right],
\end{equation}
and, as a consequence, 
\begin{equation}\label{Tuneling_esp_4}
\omega^2 \simeq \frac{V_0k^2}{4} + \mathcal O\left(1 / \cosh(4kx_0)\right). 
\end{equation}
On the other hand, also for large $kx_0 >>1$, near $x_0$, 
\begin{equation}\label{Tuneling_esp_5}
\Phi(x) \simeq V_0 \left\{\frac{1}{\cosh\left[2k(x-x_0)\right] +1} - 1 + \mathcal O\left(1/\cosh(2kx_0)\right) \right\}.
\end{equation}
Then, in a similar way to Eq.~\eqref{Tuneling_esp_4}, 
\begin{equation}\label{Tuneling_esp_6}
S_0 \simeq 2 V_0 \int_0^{x_0} dx\ \left\{\frac{3}{2} - \frac{1}{\cosh\left[2k(x-x_0)\right] +1}\right\}^{1/2} + \mathcal O(1/\cosh(2kx_0)).
\end{equation}
Then, substituting in Eq.~\eqref{Tunneling_9}, 
\begin{equation}\label{Tuneling_esp_7}
\begin{split}
E_\pm(\lambda) \simeq&  \frac{\lambda k \sqrt{V_0}}{4} \\
&\pm KT \mbox{exp}\left\{-2 \lambda V_0  \int_0^{x_0} dx\ \left\{\frac{3}{2} - \frac{1}{\cosh\left[2k(x-x_0)\right] +1}\right\}^{1/2} \right\}\\
&+ \mathcal O(1/\cosh(2kx_0)).
\end{split}
\end{equation}

\subsection{Entanglement}

Another typically quantum-mechanical phenomenon that would appear in the context of the model for quantum dynamics proposed in Sec.~\ref{sec_4} is entanglement. It must appear as long as an  asymptotic observer in Brinkmann spacetime has limited access to the whole space. That could be because of she/he only know a patch of the  whole Brinkmann space. For a small enough patch, the observer effectively lives in a lower dimensional time-like submanifold of the Brinkmann space. \\

As an example, let us to start from the ground state given in Eq.~\eqref{General_Ground_State}. For this state, the density matrix is, 
\begin{equation}\label{Entanglment_2}
\braket{v_2|\Bar \rho_{{}_\Omega}|v_1} = \frac{1}{2\pi}\int_{\mathbb R}d\lambda_1 \int_{\mathbb R} d\lambda_2 C_0(\lambda_1) C_0^*(\lambda_2) e^{i\lambda_1 v_1} e^{-i \lambda_2 v_2} \ket{0}_{\lambda_1}\bra{0}_{\lambda_2}.   
\end{equation}
Now, let us to consider  an asymptotic observer who lives in a region $v_0 - \varepsilon <v < v_0 + \varepsilon$. For small enough $\varepsilon$, the observer does not see the degrees of freedom related to the $v$-coordinate, and her/his spacetime is spanned by the coordinates $(u, q^i)$. This observer sees a reduced density matrix, 
\begin{equation}\label{Entanglement_3}
\left.\Bar \rho_{{}_\Omega}\right|_{v_0} = \int_{-\infty}^{v_0 -\varepsilon} dv\, \braket{v|\Bar \rho_{{}_\Omega}|v} + \int_{v_0 + \varepsilon}^{\infty} dv\, \braket{v|\Bar \rho_{{}_\Omega}|v}.  
\end{equation}
Strictly, this equation only satisfies for $\epsilon \rightarrow 0$, but we are going to take this limit at the end. After some computations, 
\begin{equation}\label{Entanglement_4}
\begin{split}
&\left.\Bar \rho_{{}_\Omega}\right|_{v_0} = \int_{\mathbb R} d\lambda\, \left|C_0(\lambda)\right|^2 \ket{0}_{\lambda}\bra{0}_{\lambda}\\
&{\quad}- \frac{1}{\pi}\int_{\mathbb R^2} d\lambda_1 d\lambda_2\, \frac{C_0(\lambda_1) C_0^*(\lambda_2)}{\lambda_1 - \lambda_2} e^{iv_0(\lambda_1 - \lambda_2)} \sin{\left[\varepsilon (\lambda_1- \lambda_2)\right]} \ket{0}_{\lambda_2}\bra{0}_{\lambda_1}.
\end{split}
\end{equation}
Because Eq.~\eqref{Norm_ground}, the trace of this reduced density matrix is, 
\begin{equation}\label{Entanglement_5}
\mbox{Tr}\left[\left.\Bar \rho_{{}_\Omega}\right|_{v_0}\right] = 1 - \frac{\epsilon}{\pi},  
\end{equation}
and, in the limit $\varepsilon \rightarrow 0$, we recover $\mbox{Tr}\left[\left.\Bar \rho_{{}_\Omega}\right|_{v_0}\right] = 1$, as it was expected. On the other hand, from Eq.~\eqref{Entanglement_4}, 
\begin{equation}\label{Entanglement_6}
\left.{\Bar \rho}^2_{{}_\Omega}\right|_{v_0} = \int_{\mathbb R} d\lambda\, \left|C_0(\lambda)\right|^4 \ket{0}_{\lambda}\bra{0}_{\lambda} + \mathcal O(\varepsilon^2), 
\end{equation}
and, therefore, 
\begin{equation}\label{Entanglement_7}
\mbox{Tr}\left[\left.\Bar \rho^2_{{}_\Omega}\right|_{v_0}\right]  = \int_{\mathbb R}d\lambda\, \left|C_0(\lambda)\right|^4 + \mathcal O(\varepsilon^2).
\end{equation}
Since the function $|C_0(\lambda)|^2$ must be normalized to $1$, the function $C_0(\lambda)$ satisfies:
\begin{equation}\label{Entanglement_7c}
0 < |C_0(\lambda)|^2 <1.
\end{equation} 
Therefore, in the $\varepsilon \rightarrow 0$ limit, 
\begin{equation}\label{Entanglement_7b}
\mbox{Tr}\left[\left.\Bar \rho^2_{{}_\Omega}\right|_{v_0}\right]  < \mbox{Tr}\left[\left.\Bar \rho^2_{{}_\Omega}\right|_{v_0}\right] = 1.
\end{equation}
Thus, the observer limited to live inside the region $v_0 -\varepsilon < v < v_0 + \varepsilon$, $\epsilon <<1$,  sees an entangled state coming from the pure ground state Eq.~\eqref{General_Ground_State}. This state lives at the transverse space and, in addition, the observer should measure an entangled entropy, 
\begin{equation}\label{Entanglement_9}
\begin{split}
S\left(\left.\Bar \rho_{{}_\Omega}\right|_{v_0}\right) &= -\mbox{Tr}\left[\left.\Bar \rho_{{}_\Omega}\right|_{v_0}\log \left.\Bar \rho_{{}_\Omega}\right|_{v_0} \right]  \\
&= -\int_{\mathbb R} d\lambda\, \left|C_0(\lambda)\right|^2 \log \left|C_0(\lambda)\right|^2 + \mathcal O(\varepsilon \log \varepsilon).
\end{split}
\end{equation}
Note that Eq.~\eqref{Entanglement_7}as well as Eq.~\eqref{Entanglement_9} do not depend on $v_0$. That is because the translation symmetry in the $v$ coordinate in the Brinkmann space generated by the Killing field Eq.~\eqref{CovariantlyConservedVec}. \\

As a particular example, let us to choose the ground state given by: 
\begin{equation}\label{Entanglement_1}
C_0(\lambda) =  \frac{1}{\sqrt{\pi}} e^{-\lambda^2/2}. 
\end{equation}
As exposed in Eq.~\eqref{VEV_p_v}, this choice guarantees that the pure state $\ket{\Omega}$ has zero expectation value of $\hat p_v$.  Then, 
\begin{equation}\label{Entanglement_8}
\mbox{Tr}\left[\left.\Bar \rho^2_{{}_\Omega}\right|_{v_0}\right] = \frac{1}{\pi^2}\int_{\mathbb R}d\lambda\, e^{- 2 \lambda^2} 
 + \mathcal O(\varepsilon^2)= \frac{\sqrt{\pi}}{2 \pi^2} + \mathcal O(\varepsilon^2) \approx 0.4 + \mathcal O(\varepsilon^2).  
\end{equation}
And, as expected, the observer living at $v_0 - \varepsilon< v < v_0 + \varepsilon$, $\varepsilon \ll 1$,  sees an entangled state. Substituting Eq.~\eqref{Entanglement_1} in Eq.~\eqref{Entanglement_9}, the corresponding entangled entropy is, 
\begin{equation}\label{Entanglement_10}
S\left(\left.\Bar \rho_{{}_\Omega}\right|_{v_0}\right) = - \frac{1}{\pi}\int_{\mathbb R} d\lambda\, e^{-\lambda^2}  \log \left(\frac{1}{\pi} e^{-\lambda^2}\right) + \mathcal O(\varepsilon \log \varepsilon).
\end{equation}
In the $\varepsilon \rightarrow 0$ limit, 
\begin{equation}\label{Entanglement_11}
S\left(\left.\Bar \rho_{{}_\Omega}\right|_{v_0}\right) =\frac{\sqrt{\pi}}{\pi}\left(\frac{1}{2} + \log \pi\right).
\end{equation}

\section{The field viewpoint}\label{sec_6}

In Sec.~\ref{sec_4} we set a model to build  quantum states for null particles in Brinkmann spaces and, latter, we made use of the model to study some quantum phenomena (tunneling and entanglement) in stationary Brinkmann spaces in Sec.~\ref{sec_5}. This is the quantum-mechanical point of view. However, if the reader goes back to \ref{sec_4}, she/he will realize that Eq.~\eqref{Direct_Sum_Hilbert} looks quite similar to the Fock space of some quantum field theory. In this section we explore the possibility to find a quantum field theory  for null particles in a  stationary Brinkmann space.

Let us consider a stationary Brinkmann space, with $\mathcal A_i = 0$. The reduced Hamiltonian for each value of $\lambda$ is then, 
\begin{equation}\label{Field_view_1}
H(\lambda) = \frac{1}{2}  g^{ij} p_i p_j - \frac{\lambda^2}{2} \Phi(q^k). 
\end{equation}
As we discussed in Sec.~\ref{sec_4}, $\Phi(q^k)$ must have a global maximum to guarantee the quantum stability of the system. Therefore, let us consider that $\Phi(q^k)$ has a unique maximum at $q_0^k$. Under this condition, at the low energy regime the projection of the null particles motion over  the transverse space $Q_n$ is trapped in a region near to $q_0^k$, and we can expand $H(\lambda)$ as, 
\begin{equation}\label{Field_view_2}
\begin{split}
H(\lambda) = \frac{1}{2}  g^{ij} p_i p_j -\frac{\lambda^2}{2} &\left(  \Phi_0 + \frac{1}{2} H^{(0)}_{ij} (q^i- q^i_0) (q^j- q^j_0) \right. \\
& \left. + \frac{1}{6} W^{(0)}_{ijk} (q^i- q^i_0) (q^j- q_0^j) (q^k- q^k_0) \right) + \ldots
\end{split}
\end{equation}
Where $\Phi_0 = \Phi(q^k_0)$, $H_{ij}^{(0)}$ is the Hessian at $q^k_0$ and $W_{ijk}^{(0)}$ is,
\begin{equation}\label{Field_view_3}
W_{ijk}^{(0)} = \left.\partial_i \Phi \partial_j \Phi \partial_k \Phi\right|_{q^k_0}.
\end{equation}
The contribution of $\Phi_0$ can be dropped off because it has not dynamical consequences. In addition, normal coordinates $\{\eta^i\}$ can be then introduced, at least in a neighbourhood of $q^k_0$, such that the reduced Hamiltonian $H(\lambda)$ in the low energy regime takes the form, 
\begin{equation}\label{Field_view_4}
H(\lambda) = \frac{1}{2} \delta^{ij}\Pi_i \Pi_j + \frac{\lambda^2}{2}\left( \omega_i^2 {\eta^i}^2 - \frac{1}{6} \Bar W^0_{ijk} \eta^i \eta^j \eta^k\right) + \ldots
\end{equation}
For conjugate canonical momenta $\Pi_i$ of $\eta^i$, where $- 2(\omega^i)^2$ are the eigenvalues of the Hessian $H_{ij}^{(0)}$ and  $\Bar W^{(0)}_{ijk}$ are the components of Eq.~\eqref{Field_view_3} in normal coordinates $\{\eta^i\}$. Therefore, at the low energy regime, the  reduced Hamiltonian for each $\lambda$ can be factorized as, 
\begin{equation}\label{Field_view_5}
H(\lambda) = H_{HO, \lambda}(\Pi_i, \eta^i) + H_{int, \lambda}(\eta^i),  
\end{equation}
where $H_{int, \lambda} < H_{HO, \lambda}$ in the low energy regime, for each $\lambda$, and $H_{HO, \lambda}$ is the Hamiltonian of the $n$-dimensional harmonic oscillator with  proper frequencies $\lambda \omega_i$. $H_{int,\lambda}$ will bring to an interaction term at the end of the calculus. Without this interaction term, for each $\lambda$, normal coordinates satisfy the equation of motion, 
\begin{equation}\label{Field_view_6}
\ddot \eta^i_\lambda + \lambda^2 \omega_i^2 \eta^i_\lambda = 0. 
\end{equation}
Now, if we define, 
\begin{equation}\label{Field_view_7}
\Bar \eta^i(v,t) = \int_{\mathbb R} d\lambda\ \eta^i(\lambda, t) e^{-i \lambda \omega_i v },
\end{equation}
where $\eta^i(\lambda, t) \equiv \eta_\lambda^i(t)$, Eq.~\eqref{Field_view_6} takes the form, 
\begin{equation}\label{Field_view_8}
\partial_t^2 \Bar \eta^i - \partial_v^2 \Bar \eta^i = 0,
\end{equation}
which remembers the bosonic string if we take $\eta^i$ as the coordinate fields over a world sheet  spanned by coordinates $v$ and $t$. Eq.~\eqref{Field_view_8} gives the low energy regime for null particles as viewed form the transverse space $Q_n$.  Including the interaction term $H_{int}$ in the calculus we get the first correction to the low energy regime, 
\begin{equation}\label{Field_view_9}
\partial_t^2  \Bar \eta^i - \partial_v^2  \Bar \eta^i + V^i\left(\vec{\Bar \eta}\right)= 0, 
\end{equation}
where  $V^i\left(\vec{\Bar \eta}\right)$ is an interaction term coupling the $n$ bosonic fields $\Bar \eta^i$. It is given by, 
\begin{equation}\label{Field_view_10}
V^i\left(\vec{\Bar \eta}\right) = - \frac{\delta^{il}\Bar W^{(0)}_{ljk}}{4}\int_{\mathbb R} d\lambda\ \lambda^2 \eta^j(\lambda, t) \eta^k(\lambda, t)  e^{-i \lambda \omega_i v }.
\end{equation}
because of Eq.~\eqref{Field_view_3}, this interaction term between the fields $\eta^i(v,t)$ appears provided $(\partial_i \Phi)^3|_0$ has a relevant value, \emph{i.e.} the gravitational field changes enough sharply near $q_0^k$. The next correction term would be proportional to  $(\partial_i \Phi)^4|_0$ and it would require a more suddenly change  in the gravitational field at $q_0^k$ in order to have consequences at the low energy regime.

\section{Conclusions and future work}\label{sec_7}

In this work we have proposed a way to build a quantum-mechanical model for massless particles in Brinkmann spacetimes. As a consequence of the model, quantum tunneling and entanglement of states for massless particles have been described. Also the possibility of finding a field theory at the low energy regime has been discussed. However, the massless particles has been described by means of null geodesics, without any internal degree of freedom. In this sense, the present work must be assumed as a benchmark, and the incorporation of helicity in subsequent works is a necessity to have a model approaching photons in curved spacetimes. On the other hand, generalizations to other spacetimes could be faced through the Penrose limit. In particular, the application to weak gravity in the Newtonian gauge and FLRW spacetimes could be of interest.


\begin{acknowledgments}
This work has been partially funded by 
the Escuela Polit\'ecnica Nacional under projects PIS-22-04 and PIM-19-01; 
the Ministerio Espa\~nol de Ciencia e Innovaci\'on under grant No. PID2019-107844GB-C22; the Junta de Andaluc\'ia under contract Nos. Operativo FEDER Andaluc\'ia 2014-2020 UHU-1264517, P18-FR-5057 and also PAIDI FQM-370.
\end{acknowledgments}



\bibliographystyle{ieeetr}
\bibliography{Quantum_Brinkmann}{}

\end{document}